# Redundant relationships in multiplex food sharing networks increase food security in a nutritionally precarious environment

Curtis Atkisson: Department of Anthropology, University of California, Davis

Kelly R. Finn: Animal Behavior Graduate Group, University of California, Davis; Neukom Institute, Dartmouth College

**Abstract**

Specialization is a hallmark of humans. Specialization in the real world (with imperfectly sorted partners, imperfectly calibrated supply and demand, and high failure risk) requires redundancy in relationships, which prevents specialists from going hungry when some of their partners fail to capture highly variable food items and derive the most value when dividing surplus harvests. The burgeoning field of multilayer network analysis offers tools to test for the effect of redundant relationships in food sharing networks on hunger. We derive measures that include progressively more network structure: measures without any network structure, those that only include information about individuals, and those that include all information about individuals and domains. We test for the effects of these measures in a sample of horticulturalists living in the savannahs of the Guyana Shield, a nutritionally precarious environment. Having redundant relationships is associated with a lower incidence of reported skipped meals. This provides evidence that redundancy in food sharing networks may mitigate risk associated with the foraging strategies necessary to support a large-brained, generalist omnivore. This result has consequences for broader debates in the field of human evolution such as why humans live in groups with low relatedness.

**Introduction**

Specialization is a hallmark of the human species and may have played a key role in the evolution of human intelligence (Kaplan et al., 2000). Humans require a diversity of macro- and micro-nutrients to stay healthy and dense resource packages to grow and maintain our large brains (Aiello and Wheeler, 1995). Such packages are often rare (e.g., large game), variable (e.g., a loss of a crop), or difficult to use (e.g., digging and processing bitter cassava). As large-brained omnivores, only with trade between specialized resource producers could humans acquire both the dense and diverse nutrients to successfully occupy the highly productive niche that was key to the evolution of our distinct life history traits (Gurven, 2004). Successful specialization, then, decreases risk of hunger while maintaining a sufficiently varied diet.

Recent research shows that there are large and consistent differences in harvests between individuals, which may be attributable to differences in skill (McElreath and Koster, 2014; Koster et al., 2019; also Marshall, 1976; Kaplan and Hill, 1985; and Hawkes, 1993). If this is the case, we would expect individuals to specialize in foraging strategies in which they are most skilled (Durkheim, 1893; Winterhalder, 1996). This would lead them to overproduce the thing they harvest and frequently give it away, but also to need partners who give them the items in which they do not specialize to maintain diet diversity (Ricardo, 1817; Schwartz and Hoeksema,1998).

Specialization as a successful strategy is therefore contingent on the relationships between specialists, as much as it is on ability to specialize. It is not only necessary that partners get resources at a high enough frequency for themselves, but also that they provide the resource to their partners. Having multiple partners one can exchange the same resources to and from (i.e. redundant trade relationships) means individuals have a higher chance of having a sufficient and sufficiently varied diet without waste.

We use multilayer network representations of trade patterns to characterize both what individuals trade, and with whom they trade. Multilayer networks are comprised of multiple connected layers of networks, and multiplex networks are special cases where nodes in different layers only connect to versions of themselves, often as layers of the same networks in different domains of social interaction (Kivelä et al., 2014). While this concept of 'multiplexity' is not new (Mitchell, 1969; Hinde, 1978), the tools to analyze multi-layer social networks are only now being developed (e.g., De Bacco, et al., 2017), and calls have been issued for their use in ecology (Pilosof et al, 2018), animal behavior (Finn et al., 2018), and human behavioral evolution (Atkisson et al., 2020). Multiplex networks provide excellent means to measure patterns of food sharing because they retain detail about both who one trades with and in which domains two individuals trade. This approach therefore allows for characterization and measurement of trade patterns, including unique or redundant trade relationships.

We assess the relationship between trade patterns of individuals and their food security in a horticultural population subject to high variance in both hunting and farming returns. We quantify an individual's pattern of food sharing with measures of their multiplex food sharing networks that capture variation and redundancy in what they trade and with whom. One class of measures consists of summary statistics for each individual that are independent of network structure. A second class of measures is derived from collapsing all layers of the multiplex network onto a single aggregate network, which distinguishes trade amongst specific partners but not about specific items of trade. The third class of measures is derived from the full multiplex network, which distinguish different trade partners and items. We examine the empirical fits of models that predict household food security with different sets of these measures. These results allow us to characterize the structure of an individual's food sharing relationships and explore which aspects of their network help mitigate risk of hunger amongst a population in a nutritionally precarious environment.

**Methods**

*Makushi of Guyana*

Data were collected from 270 participants across nine villages of the North Rupununi region of Guyana, which is primarily populated by Makushi people. Guyanese Makushi, are relatively isolated with little infrastructure, leading them to use traditional subsistence techniques. The Makushi are swidden horticulturalists who primarily plant cassava (*Manihot esculenta*; Schacht, 2014). They supplement their horticulture with protein sources that shift over the year according to the season: in the dry season (8-9 months of the year) fish and in the rainy season (3-4 months of the year) wild meat. They also supplement their diet with purchased sugar, rice, beef, and frozen chicken.

*Social network data*

Within each village, a census of households was conducted, and 30 households were selected at random, with one adult interviewed. Respondents were asked questions about giving and receiving eight food types selected to cover the range of ways people can get food. These include parched cassava, cassava bread, and rice as the main carbohydrates; wild meat, fish, and purchased meat as the various protein sources; and oil and sugar as important purchased ingredients. These data were gathered using a culturally specific name generator (Wasserman and Faust, 1984). A multiplex ego-network was constructed from these data for each person—a multiplex network because an individual and their trade partners are included in each layer with the presence or absence of interactions, and an ego-network because it only represented an individual and their ties (Wasserman and Faust, 1984). Note that the measures in this study do not include the amount traded or frequency of trade, just presence of trading relationships that are unique by trade partner, direction of trade, and food source.

*Outcome variables*

We examine the effect the structure of the multiplex networks has on food security. The two outcomes we examine are how often in a week someone in the house skips a meal because there is no food (mean=1.30, sd=1.17) and how often in a week someone goes to bed hungry because there is no food (mean=1.17, sd=1.08). These variables are a count of days and are modeled as a negative binomial.

*Measures*

We use three sets of predictor variables derived from trade patterns calculated (1) independent of network structure, (2) from a single layer aggregated network, and (3) from the multiplex networks. They are described in detail below and are given colloquial reference in Table 1. Figure 1 shows examples of these variables calculated on two hypothetical networks.

| Table 1: Definition of measures | | |
|---|---|---|
| *Measure* | *Description* | *What it means when maximized* |
| Partners | # of partners | Trades with many people |
| Domains | # of domains (max = 16) | Gives and receives in many domains |
| Strength | # of partner-domain interactions | Interacts with many people in many domains |
| Importance diversity | The diversity of import of partners | All partners are equally important |
| N-gram diversity | The diversity of trade relationships | All trade relationships are unique |
| Redundancy | Amount of redundancy for a given diversity of import of partners | There are few trade relationships that have many partners |

**Figure 1: An example of multiplex and aggregate networks and measures calculated on them**

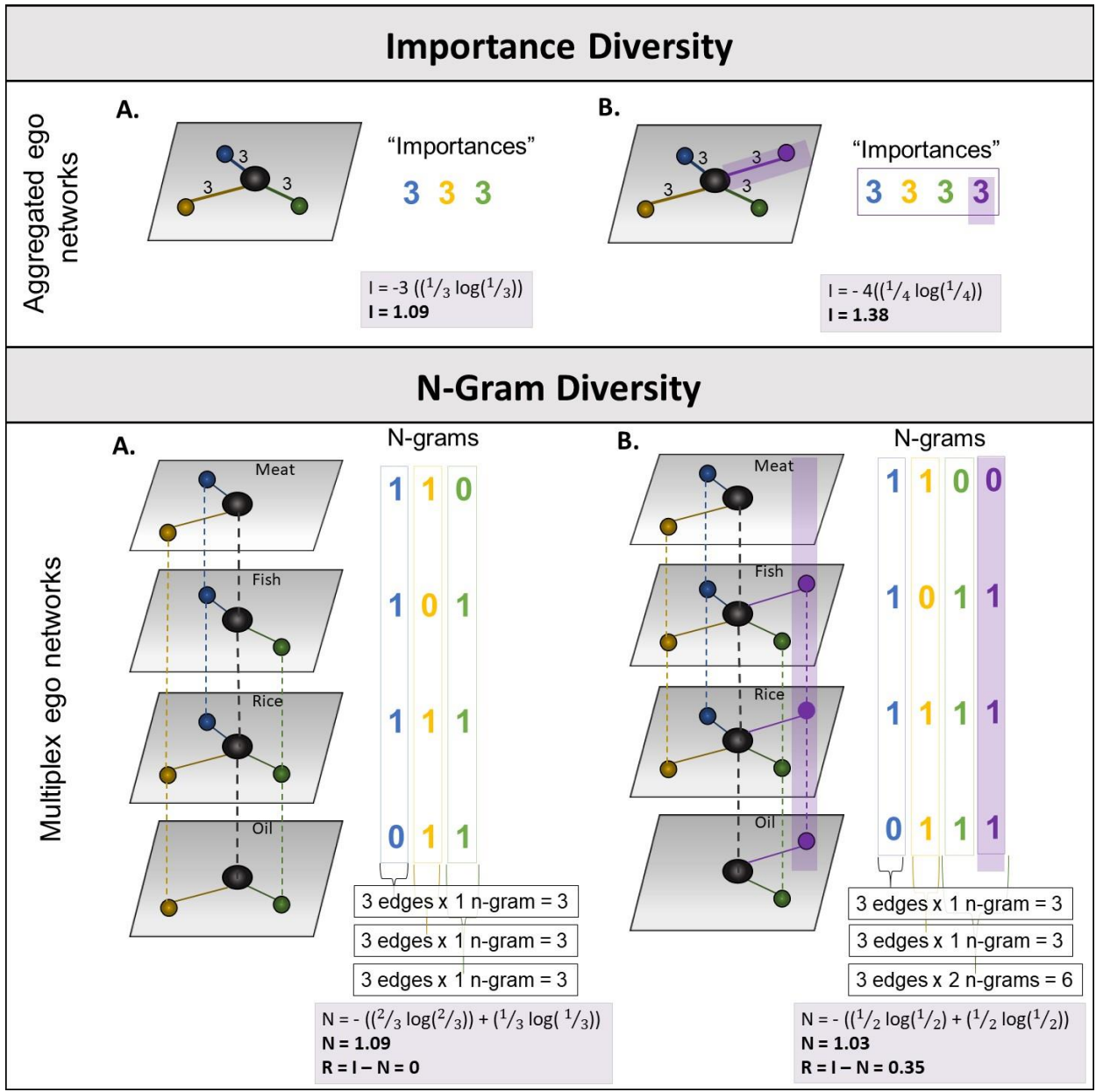

*Figure 1: This figure shows two hypothetical multiplex networks (A and B). There are three partners in A (green, yellow, and blue) and four partners in B (as before with purple). All three of the partners in A have a unique relationship type, but each of the relationships has three connections in it. This leads to a calculation of the N-gram diversity on the vector {3,3,3}, which equals 1.09. In network B, two individuals (green and purple) have the same relationship type. This leads to a calculation of the N-gram diversity on the vector {3,3,6}, which equals 1.03. Meanwhile, all individuals have the same importance on both networks, we have just added one individual to network B. Their respective vectors upon which entropy is calculated are {3,3,3}=1.09 and {3,3,3,3}=1.38. By taking the difference of Importance and N-gram diversity,*

*we get a measure of the redundancy in each multiplex network. Network A has a redundancy of R=0 while network B has a redundancy of R=0.35.*

All measures on the aggregate or multiplex networks use information theory to quantify diversity (i.e. Shannon-Weiner index). Please see Supplementary materials for a summary of information theory (*S1*) and the detailed equations for each measure (*S2*).

<u>Measures without network structure</u> Some basic features of an individual's social networks can be indexed as summary statistics of frequencies of trade behaviors, without even using social network structure to distinguish between what they trade with whom. These include the number of *domains* (Kasper and Borgerhoff Mulder, 2015) in which they have partners, the number of *partners* they have, and the sum of binary values (0/1) reflecting if they have interactions with a partner in a domain (*strength*).

<u>Aggregate-layer measure</u> Interactions across layers can be collapsed into a single-layer aggregate network where the weight of the tie between two individuals is the sum of binary edges in all domains in which they interact (see top panel in Figure 1). These edge weights are then divided by the total strength as 'importances' of relationships (i.e. the proportion of unique trades coming from a partner). If an individual gives and receives many different food items from the same partner, that relationship has high "importance". The Shannon entropy of these values gives a measure of diversity of importances across relationships that we refer to as *importance diversity* (a modification of "Partner diversity" in Silk et al., 2013). Individuals with high importance diversity have trade interactions distributed evenly across many partners. In other words, each trade event is equally likely to occur with any trade source. If a high proportion of trades occur with one or a few of their partners, the diversity of their trades across partners is lower. Thus, importance diversity decreases as an individual has fewer partners, and/or as they trade in disproportionately more domains with some partners. It reaches zero if they only trade with one partner. See equation S2: *I*.

<u>Multiplex measures</u> Using the full multiplex ego network, each individual-partner pair has a specific pattern of connections across domains (see bottom panel of Figure 1). Each relationship is represented as an N-gram, a string of N number of 0's and 1's, indicating whether they interacted in N=16 domains. Relationships with two different partners are considered the same "type" if they show the same patterning of trade (see purple and green nodes in network B in Figure 1). The value for each relationship type is the total number of interactions in that relationship type (3 in Fig 1) multiplied by the number of partners who have that same type of relationship. This value (*N* in equation S1), *N-gram diversity*, indexes how trade is distributed across relationship types—it will be maximized when all partners have a unique relationship type with the same amount of trade. Like importance entropy, it decreases as more trade occurs between certain partners, but it also decreases when partners share a single relationship type.

When each partner has a unique relationship type, *N-gram* and *importance* diversity are the same (see network A in Figure 1). Any duplicate N-gram decreases the N-gram diversity relative to the importance diversity. As such, we also calculate the difference between these two and call it *N-gram redundancy* (*R* in Fig 2). A large difference between these two values means that an individual has many partners that share the same pattern of interactions. See *Figure S2* for a comparison of N-gram and importance diversity.

*Analysis plan*

The relationship between the outcome and independent variables is explored using negative binomial regressions. The effects are estimated in a Bayesian way using Monte Carlo Markov Chains (MCMC) with a No U-Turn Sampler (NUTS) as implemented in STAN (Carpenter, et al., 2017). The package brms (Bürkner, 2017) is used for convenience in R (R Core Team).

To examine if and how redundancy in food sharing networks mitigates risk, we test four models for each outcome variable. The first model includes only the measures without network structure: number of partners, number of domains, and total strength. The second model includes the measures without network structure as well as the measure on the single-layer aggregate network: importance diversity. The third model includes the measures without network structure and N-gram redundancy. The fourth model includes the measures without network structure, importance diversity, and N-gram redundancy. Models including measures without network structure with N-gram diversity, and including measures without network structure, importance diversity, and N-gram diversity are presented in the supplementary material (*Table S1*). If a model including N-gram redundancy has the most predictive power, it would suggest the structure of relationships across domains is an important aspect of how trade is patterned, as it relates to hunger outcomes. We use leave-one-out (loo) cross-validation information criteria (loo_ic) to judge models (Stone, 1977). From this we are able to get a mean and standard error of the difference between models. If the standard error is equal to or larger than the mean, we have little confidence in our estimate of the difference between models.

**Results**

**Table 2: Estimated effect of measures derived from the multiplex network in 4 models of 2 outcomes**

| Outcome | Model ID | No-network-structure | | | | | | Aggregate-layer | | Multiplex | | loo_ic |
|---|---|---|---|---|---|---|---|---|---|---|---|---|
| | | *Strength* | *Error* | *Domains* | *Error* | *Alters* | *Error* | *Importance* | *Error* | *Redundancy* | *Error* | |
| **Skip** | N | -0.01 | 0.01 | 0.05 | 0.02 | 0.01 | 0.02 | ---- | ---- | ---- | ---- | 753.19 |
| | N + A | -0.01 | 0.01 | 0.05 | 0.02 | 0.01 | 0.04 | -0.02 | 0.22 | ---- | ---- | 755.56 |
| | N + M | 0.01 | 0.01 | 0.03 | 0.02 | 0.01 | 0.02 | ---- | ---- | -0.71 | 0.32 | 750.05 |
| | N + A + M | 0.01 | 0.01 | 0.03 | 0.03 | 0.01 | 0.04 | 0.02 | 0.23 | -0.71 | 0.32 | 752.88 |
| **Hungry** | N | -0.01 | 0.01 | 0.07 | 0.02 | -0.01 | 0.02 | ---- | ---- | ---- | ---- | 707.55 |
| | N + A | -0.01 | 0.01 | 0.07 | 0.02 | -0.01 | 0.04 | 0.03 | 0.24 | ---- | ---- | 709.49 |
| | N + M | -0.01 | 0.01 | 0.06 | 0.03 | 0.01 | 0.02 | ---- | ---- | -0.30 | 0.32 | 709.15 |
| | N + A + M | -0.01 | 0.01 | 0.06 | 0.03 | -0.01 | 0.04 | 0.06 | 0.24 | -0.30 | 0.32 | 711.45 |

The results for the four models for both outcome variables are shown in Table 2. When only measures with no network structure are included (Model N), there is no clear relationship between them and the outcomes, aside from the number of domains in which an individual shares food (effect estimates for *Strength* and *Partners* are not larger than their errors, while *Domains* is only a little bigger). When we add *importance diversity* to the analysis (Model N+A), no predictor variables have clear relationships to the outcome variables, with the same exception. Only when adding *N-gram redundancy* (Models N+M and N+A+M) to the analysis is anything a clear, meaningful predictor (though in the model of hungry, we have lower confidence in its importance). This provides support that redundancies in relationship types predicts increased food security.

Model IDs correspond to Model IDs in Table 3.

We estimated the marginal effect from the best model of N-gram redundancy on how many days in a week someone skips a meal (Figure 2). See Supplementary materials *S3* for evidence that this relationship is not driven by extreme values of redundancy.

**Figure 2: Marginal effect of multiplex structure on skipping a meal**

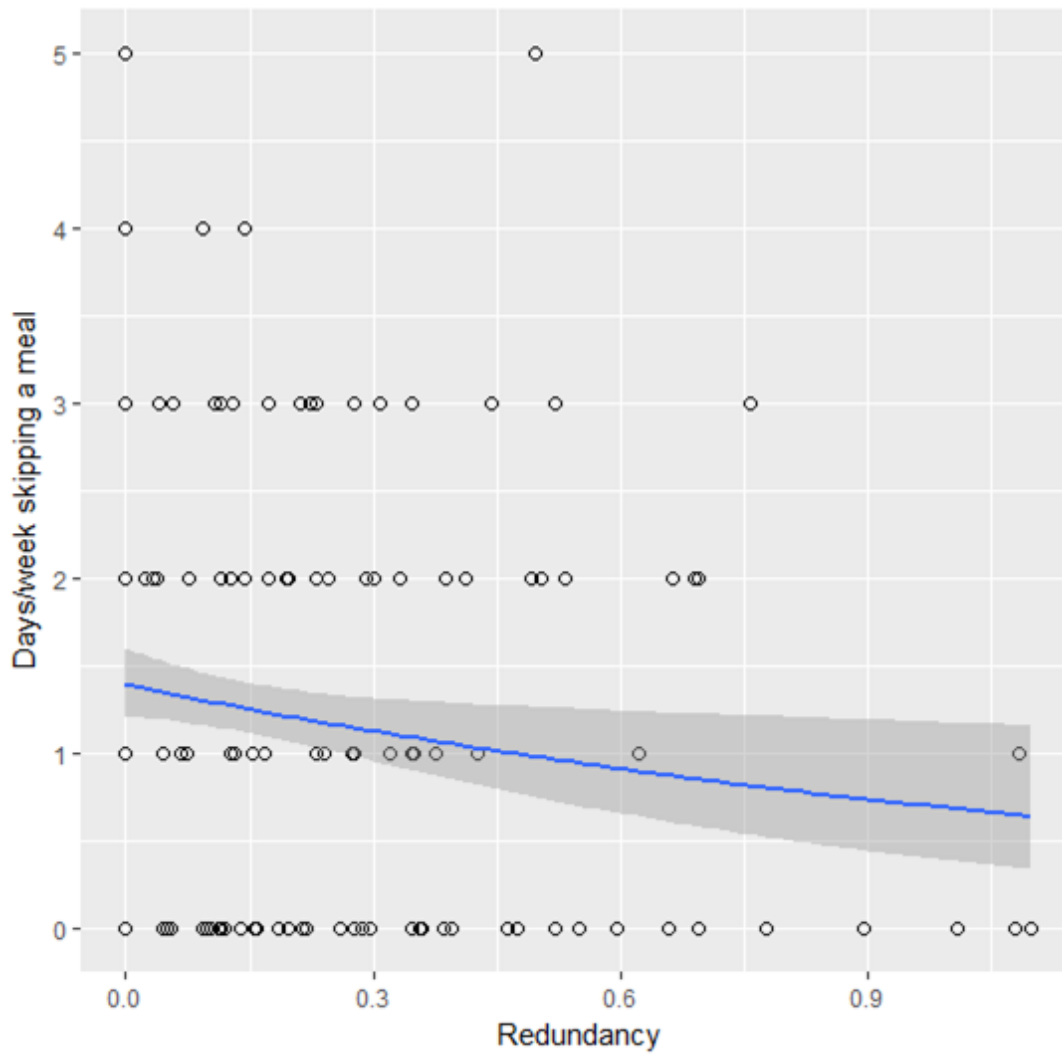

*Figure 2: Marginal effect of N-gram redundancy (a measure of redundancy in food sharing relationships) on the number of days per week an individual goes hungry. The blue line shows the mean and the shaded areas the 95% credibility interval. The points are the original data.*

For models that predict skipping a meal, three model comparisons fulfill the criterion of having a larger mean difference than standard error of the difference (all comparisons can be found in *Table S2*). We have confidence that the model with measures without network structure is better than the model with measures without network structure and the aggregate network measure (N – [N + A] = -2.36, std. error = 0.52); the model with measures without network structure and the multiplex network structure measure is better than the model with measures without network structure and the aggregate network measure ([N + M] – [N + A] = -5.51, std. error = 4.87); and the model with measures without network structure and the multiplex network structure measure is better than the full model ([N + M] – [N + A + M] = -2.83, std. error = 0.62). Given this, even though loo_ic provides no direct support for N + M being better than all alternative models, we have confidence that it is the best model.

For number of days per week one goes to bed hungry, however, the effect of multiplex structure is not well-supported. It seems clear that the best model would not include the aggregate-network measure but less clear if a best model would also include the multiplex network measure or not.

**Discussion**

These results show that individuals with redundant relationships—showing the same patterning of trade items with multiple trade partners—have greater food security. This redundancy likely acts as a buffer to risk in the nutritionally poor environment occupied by the Makushi and could generalize to other subsistence groups occupying neo-tropical savannahs (e.g., the Hiwi, see Hurtado and Hill, 1990).

Specializing in a subset of all required resources likely decreases variability in harvesting those resources. For example, Makushi villages lie in the middle of the savannah, several hours walking from hunting grounds that are good in the dry season. Someone who specializes in hunting, therefore, will have a hunting trip of more than 6 hours for small prey (e.g., large rodents such as agouti) or multiple days for large prey (e.g. deer). This person must become an expert in finding and following trails and devote large amounts of time to hunting. By specializing, the skilled hunter can learn to more efficiently capture prey. However, this leaves no time for wage labor, working a farm, or daily fishing.

Such specialists, therefore, will often trade their surplus of resources for the remaining resources they need, for which they are not specialized. Our data provide evidence that the redundancy in relationship types, but not other types of redundancy (e.g., in total trading interactions or total partners), leads to increased food security. People in this population who have established multiple relationships with the same pattern of giving and receiving resources experience greater food security, which is what we would expect of specialists.

While redundancy of relationship type did predict skipping meals, it was not a well-supported predictor of going to sleep hungry. Ethnographic experience indicates that people at this site are averse to going to sleep hungry. As such, people may labor all day with no food simply to have something to eat at night, in which case they would skip meals but not go to sleep hungry leading to a decreased frequency and variability in going to sleep hungry, which is what is found here. This type of decision making could potentially explain the difference in these food security outcome variables.

Our data also show that the number of domains in which an individual traded was predictive of skipping meals more often. This effect size was small compared to that of redundancy, (predicting less than a half day increase in expected number of days spent hungry across the entire range of number of domains in contrast to nearly one day decrease across the entire range of redundancy), but consistently supported. Increases in the total domains in which an individual interacts are primarily driven by giving in more domains than expected (see Supplementary materials *S4*). Individuals, therefore, are transferring food out even though doing so decreases their food security. This might indicate that food is being exchanged for a different resource or even that these people are less specialized.

These results that suggest foraging risk is mitigated by redundant trade relationships have broader implications for human evolution. Cooperative breeding is a primary explanation for the rise of human sociality and our pattern of life history traits (Hawkes et al., 1998). It is easy to see how cooperative breeding may arise in the context of groups with high average relatedness, yet analyses indicate that humans often live in groups where the average relatedness is low (Hill et al., 2011). However, given a diverse diet of items that require skill to extract, if the skills for extracting certain resources are vertically transmitted either through genetics or learning (i.e. members of the same family are likely to specialize in the same domains), then people would need to live in groups with low average relatedness in order to acquire a diverse diet and have redundancy in their trade networks. Tolerating non-kin and sharing food items that require skill to harvest may have been key for humans to occupy the niche that allowed for the evolution of large brains.

Vertical transmission of skill and trade across families can indeed be observed in this population. Traditional Makushi farms have over 60 varieties of cassava with different characteristics (e.g., grow well in rainy conditions) and different preparation requirements. Some families specialize in growing cassava and have extensive networks for sharing cassava varieties across a large range. A family may propagate some cassava varieties, send people to trade for the varieties they do not propagate, and carefully maintain the growing crop. This requires a large amount and variety of work, and often the entire family will devote a large portion of their time to cassava farming and trade. This system massively increases stability in cassava harvests but leaves little time for other efforts. However, sharing their casava harvest with families who specialize in other things, they can get both the quantity and diversity of resources they need.

This work creates a framework to mathematically assess an individual's relationships across unique individuals and unique social domains and demonstrates how measures can be developed to capture specific aspects of an individual's social situation. Of the various network measures we derived, the measure of relationship redundancy that requires preserving the multiplex network structure (i.e. both partners and domains) best predicted hunger risk, linking trade patterns to risk mitigation. This relationship would not be identified with more general questions about trade (e.g., "With whom did you share food in the last month?"). Future studies on food sharing should gather information about food sharing in more than one domain and focus on an individual's pattering of trade across their relationships. We predict that such studies will find that redundancy is more important in environments with highly variable foraging activities, especially in environmentally precarious environments.

Many anthropologists gathering social network data currently gather multiplex data. Multiplex networks can be constructed when individuals get asked about the same group, when all individuals *could be* present in all other individuals' networks, and when there are multiple types of interactions or relationships (even kinship could be its own network). Despite often collecting data at this level of detail, it is common to collapse data into measures without network structure (e.g., number of partners) or aggregate-network measures (e.g., through summing total number of relationships and calculating a network measure such as centrality). Sometimes this may be sufficient learn about the dynamics underlying our data, but it may also throw away valuable information that could help us better understand our data (Atkisson, et al., 2020). To preserve this additional structure, data can be represented as complete multiplex networks when every person in a community is asked the same set of questions, or as multiplex ego-networks, such as in this paper. We encourage people with data of this type to consider multi-layer analyses.

**Acknowledgements**

We would like to thank Monique Borgerhoff Mulder, Cristina Moya, and Andy Sih for extensive comments on previous drafts. The EEHBC lab at University of California, Davis gave invaluable advice. Hon. Ryan James offered helpful conversation about information theory. All remaining mistakes are our own. Sydney Allicock was invaluable for his approval of the project through the Ministry of Amerindian Affairs (now the Ministry of Indigenous Peoples' Affairs) in Guyana. Rebecca Xavier and Ricky Moses gathered data in the field. We are especially grateful for the warmth and easy participation of the people of the North Rupununi area of Guyana.

**Funding**

Funding for data collection for this research came from the National Science Foundation (# 1558890) and The Wenner-Gren Foundation (Dissertation Fieldwork Grant). This work was further supported by a University of California, Davis Provost's Dissertation Year Fellowship awarded to CA. KRF was supported by the U.S. Army Research Office MURI Award No. W911NF-13-1-0340, discretionary funds from PhD advisor Brenda McCowan, and the Neukom Institute at Dartmouth College.

**References**

Kaplan, H., Hill, K., Lancaster, J., & Hurtado, A. M. (2000). A theory of human life history evolution: Diet, intelligence, and longevity. *Evolutionary Anthropology*, *9*(4), 156–185. https://doi.org/10.1002/1520-6505(2000)9:4<156::AID-EVAN5>3.0.CO;2-7

Gurven, M. (2004). To give and to give not: The behavioral ecology of human food transfers. *Behavioral and Brain Sciences*, *27*(4), 543–559. https://doi.org/10.1017/s0140525x04000123

Jaeggi, A. V., Hooper, P. L., Beheim, B. A., Kaplan, H., & Gurven, M. (2016). Reciprocal Exchange Patterned by Market Forces Helps Explain Cooperation in a Small-Scale Society. *Current Biology*, *26*(16), 2180–2187. https://doi.org/10.1016/j.cub.2016.06.019

Kasper, C., & Mulder, M. B. (2015). Who helps and why?: Cooperative networks in Mpimbwe. *Current Anthropology*, *56*(5), 701–732. https://doi.org/10.1086/683024

Trivers, R. (1971). The Evolution of Reciprocal Altruism. *The Quarterly Review of Biology*, *46*(1). https://doi.org/10.1086/406755

Wilkinson, G. S. (1984). Reciprocal food sharing in the vampire bat. *Nature*, *308*(5955), 181–184. https://doi.org/10.1038/308181a0

Carter, G. G., & Wilkinson, G. S. (2013). Food sharing in vampire bats: Reciprocal help predicts donations more than relatedness or harassment. *Proceedings of the Royal Society B: Biological Sciences*, *280*(1753). https://doi.org/10.1098/rspb.2012.2573

De Waal, F. B. M. (1997). The chimpanzee's service economy: Food for grooming. *Evolution and Human Behavior*, *18*(6), 375–386. https://doi.org/10.1016/S1090-5138(97)00085-8

De Waal, F. B. M. (2000). Attitudinal reciprocity in food sharing among brown capuchin monkeys. *Animal Behaviour*, *60*(2), 253–261. https://doi.org/10.1006/anbe.2000.1471

Hauser, M. D., Chen, M. K., Chen, F., & Chuang, E. (2003). Give unto others: Genetically unrelated cotton-top tamarin monkeys preferentially give food to those who altruistically give food back. *Proceedings of the Royal Society B: Biological Sciences*, *270*(1531), 2363–2370. https://doi.org/10.1098/rspb.2003.2509

Rutte, C., & Taborsky, M. (2008). The influence of social experience on cooperative behaviour of rats (Rattus norvegicus): Direct vs generalised reciprocity. *Behavioral Ecology and Sociobiology*, *62*(4), 499–505. https://doi.org/10.1007/s00265-007-0474-3

McElreath, R., & Koster, J. (2014). Using Multilevel Models to Estimate Variation in Foraging Returns: Effects of Failure Rate, Harvest Size, Age, and Individual Heterogeneity. *Human Nature*, *25*(1), 100–120. https://doi.org/10.1007/s12110-014-9193-4

Koster, J., McElreath, R., Hill, K., Yu, D., Shepard, G., van Vliet, N., … Ross, C. (2019). The Life History of Human Foraging: Cross-Cultural and Individual Variation. *BioRxiv*, 574483. https://doi.org/10.1101/574483

Durkheim, E. (1893). *The Division of Labor in Society*. New York: Free Press.

Winterhalder, B. (1996). Social foraging and the behavioral ecology of intragroup resource transfers. *Evolutionary Anthropology*, *5*(2), 46–57. https://doi.org/10.1002/(SICI)1520-6505(1996)5:2<46::AID-EVAN4>3.0.CO;2-U

Kivelä, M., Arenas, A., Barthelemy, M., Gleeson, J. P., Moreno, Y., & Porter, M. A. (2014). Multilayer networks. *Journal of Complex Networks*, *2*(3), 203–271. https://doi.org/10.1093/comnet/cnu016

Mitchell, J. C. (1969). The concept and use of social networks. In J. C. Mitchell (Ed.), *Social networks in urban situations*. The University Press.

Hinde, R. A. (1978). Interpersonal relationships - in quest of a science. *Psychological Medicine*, *8*(3), 373–386. https://doi.org/10.1017/S0033291700016056

De Bacco, C., Power, E. A., Larremore, D. B., & Moore, C. (2017). Comdmunity detection, link prediction, and layer interdependence in multilayer networks. *Physical Review E*, *95*(4), 1–10. https://doi.org/10.1103/PhysRevE.95.042317

Finn, K. R., Silk, M. J., Porter, M. A., & Pinter-Wollman, N. (2019, March 1). The use of multilayer network analysis in animal behaviour. *Animal Behaviour*. Academic Press. https://doi.org/10.1016/j.anbehav.2018.12.016

Atkisson, C., Górski, P. J., Jackson, M. O., Hołyst, J. A., & D'Souza, R. M. (2020). Why understanding multiplex social network structuring processes will help us better understand the evolution of human behavior. *Evolutionary Anthropology*.

Torrado, M. (2007). *Road Construction and Makushi Communities of Southern Guyana: Impacts and Consequences*. Syracuse University.

Jaynes, E. T. (1957). Information theory and statistical mechanics. *Physical Review*, *106*(4), 620–630. https://doi.org/10.1016/b978-008044494-9/50005-6

Shannon, C. E., & Weaver, W. (1963). *The mathematical theory of communication. The University of Illinois Press*. Champaign-Urbana. https://doi.org/10.1145/584091.584093

Carpenter, B., Gelman, A., Hoffman, M. D., Lee, D., Goodrich, B., Betancourt, M., … Riddell, A. (2017). Stan: A probabilistic programming language. *Journal of Statistical Software*, *76*(1).

Bürkner, P.-C. (2017). brms: An R Package for Bayesian Multilevel Models Using Stan. *Journal of Statistical Software*, *80*(1).

R Core Team. (2017). R: A language and environment for statistical computing. Vienna, Austria: R Foundation for Statistical Computing. Retrieved from https://www.r-project.org/

Stone, M. (1977). An Asymptotic Equivalence of Choice of Model by Cross-Validation and Akaike 's criterion. *Journal of the Royal Statistical Society*, *39*(1), 44–47.

Vehtari, A., Gelman, A., & Gabry, J. (2017). Practical Bayesian model evaluation using leave-one-out cross-validation and WAIC. *Statistics and Computing*, *27*(5), 1413–1432. https://doi.org/10.1007/s11222-016-9696-4

Hawkes, K., O'connell, J. F., Blurton Jones, N. G., Alvarez, H., & Charnov, E. L. (1998). *Grandmothering, menopause, and the evolution of human life histories* (Vol. 95). Retrieved from www.pnas.org.

Hill, K. R., Walker, R. S., Božičević, M., Eder, J., Headland, T., Hewlett, B., … Wood, B. (2011). *Co-Residence Patterns in Hunter-Gatherer Societies Show Unique Human Social Structure*. Retrieved from http://science.sciencemag.org/

Schwartz, M. W., & Hoeksema, J. D. (1998). Specialization and Resource Trade: Biological Markets as a Model of Mutualisms. *Ecology*, *79*(3), 1029. https://doi.org/10.2307/176598

Ricardo, D. (1817). *On The Principles of Political Economy and Taxation*. London: John Murray.

Hammerstein, P., & Noë, R. (2016). Biological trade and markets. *Philosophical Transactions of the Royal Society B: Biological Sciences*, *371*(1687). https://doi.org/10.1098/rstb.2015.0101

Kaplan, H. S., & Hill, K. R. (1985). Food Sharing Among Ache Foragers: Tests of Explanatory Hypotheses. *Current Anthropology*, *26*(2), 223–246.

Hawkes, K. (1993). Why hunter gatherers work: An ancient version of the problem of public goods. *Current Anthropology*, *34*(4), 341–361. Retrieved from http://www.jstor.org/stable/2743748

Marshall, L. (1976). Sharing, talking, and giving: Relief of social tensions among !Kung Bushmen. In R. B. Lee & I. DeVore (Eds.), *Kalahari hunter gatherers: Studies of the !Kung San and their neighbors*. Cambridge: Harvard University Press.

Silk, J., Cheney, D., & Seyfarth, R. (2013). A Practical Guide to the Study of Social Relationships, *22*, 213–225. https://doi.org/10.1002/evan.21367

Schacht, R. N. (2015). Cassava and the Makushi : A Shared History of Resiliency and Transformation Cassava and the Makushi : A Shared History of Resiliency and Transformation, (April). https://doi.org/10.5040/9781350042162.ch-001

**Ethics**

This study was approved by the Institutional Review Board at the University of California, Davis (IRB #842779-2), The Ministry for Amerindian Affairs (now the Ministry of Indigenous Peoples' Affairs) of Guyana, The North Rupununi District Development Board, and the leadership of each participating community.

**Data, code and materials**

Finalized code will be made available via github

**Competing interests**

There are no competing interests

**Authors' contributions**

CA conceived of, designed, and coordinated the study, gathered data, analyzed the data, created the measures, created figures, and drafted the paper; KA created the measures, created figures, and critically revised the manuscript. All authors gave final approval for publication and agree to be held accountable for the work performed therein.

## Supplementary Materials

*S1: Information theory and diversity*

We use information theoretic approaches for quantifying the patterning of trade across partners and domains of an individual's food sharing ego network. Information theory has developed tools to quantify the patterning or 'information content' in each assemblage of data (Jaynes, 1957). Shannon entropy, the most foundational of such tools, measures what could partnernatively be called disorder, diversity, mixedupness, and unpredictability of a system using a universal unit called 'bits' (Shannon and Weaver, 1963). It is calculated from the probabilities that 'events' come from one of many 'sources' (i.e. the probabilities of different event types). A system has maximum entropy if 'events' are equally likely to come from a number of different 'sources' (i.e. all event types are equally as likely); it is maximally disordered, diverse, mixed-up, and unpredictable such that more 'bits of information' are needed to describe the system. A system has minimum or no entropy if all 'events' come from the same 'source' (i.e. only one event type will happen); it is completely ordered, uniform, sorted, and predictable and no information is needed to describe its patterning.

Entropy is calculated as

$$-\sum_{i=1}^{n} p_i log(p_i) \tag{S1}$$

where *p* is the frequency of events from each category *i*. When all occurrences are from the same category ($p_i = 1$), entropy will be 0. As the items get spread between different categories, this value will increase. One of the ways of interpreting entropy is that the larger the number, the more diverse the origin of the items in the assemblage is. Measures derived from information theory are ideal for quantifying patterning of food sharing in a social network because they can be designed to reflect redundancy in different aspects of trade, as described in the measures below.

*S2: Measures*

<u>Importance diversity</u> – Importance diversity is defined as

$$I = -\sum_{j=1}^{J} \left( \frac{\sum_{d=1}^{D} a_{dj}}{\sum_{d=1}^{D} \sum_{j=1}^{J} a_{dj}} log \left( \frac{\sum_{d=1}^{D} a_{dj}}{\sum_{d=1}^{D} \sum_{j=1}^{J} a_{dj}} \right) \right) \tag{S2}$$

Where $a_{dj}$ is a tie on domain *d* with individual *j*, *j* indexes each partner, *J* is the total number of partners, *d* indexes each domain, and *D* is the total number of domains.

<u>N-gram diversity</u> – N-gram diversity is defined here. If we consider $z_d$ as a relationship type (an N-gram; a specific pattern of 1's and 0's denoting connections on each domain), then

$$Z = \sum_{d=1}^{D} z_d \tag{3}$$

is a positive integer and the weight of the N-gram. *N-gram diversity* is defined as

$$N = -\sum_{j=1}^{J}\left(\frac{\left(\sum_{d=1}^{D}\begin{cases}1 & if\, a_j = z_d \\ 0 & if\, a_j \neq z_d\end{cases}\right)Z}{\sum_{d=1}^{D}\sum_{j=1}^{J}a_{dj}} log\left(\frac{\left(\sum_{d=1}^{D}\begin{cases}1 & if\, a_j = z_d \\ 0 & if\, a_j \neq z_d\end{cases}\right)Z}{\sum_{d=1}^{D}\sum_{j=1}^{J}a_{dj}}\right)\right)$$

(4)

where *a*, *d*, *D*, *j*, and *J* are as before.

*S3: N-gram redundancy and skipping a meal*

Fig S1 shows a violin plot from the data for N-gram redundancy by number of days per week an individual skips a meal. If the effect was being driven only by a few points (e.g., all individuals above .5 redundancy never skip meals), we would see similar centers of mass for each violin. Instead, the center of mass shifts and the tail gets longer for each violin. This indicates that the effect is across the entire range of redundancy, not simply driven by a few points. Furthermore, there were no pareto-k values that indicated undue influence by a single datapoint (Vehtari, et al., 2017).

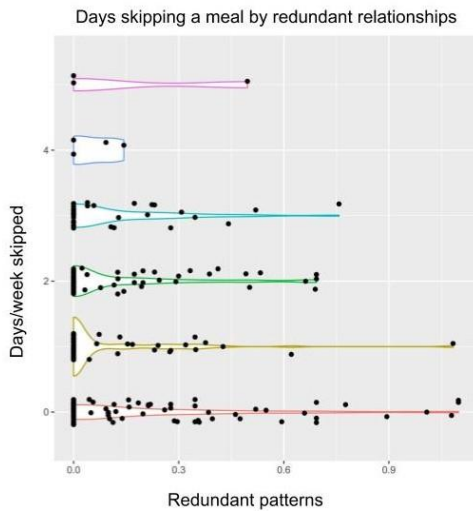

*S4: Increased total number of domains*

As reported in the main text, the number of domains is a consistently supported predictor of decreased food security. By comparing the number of domains in each direction (giving and receiving) we can see which direction of domain is driving increases in the total number of domains. To do so, we take the total number of domains in which an individual has interactions and divide that by 2, representing a naïve expectation of half of a person's domains going each direction. There are a total of 82.5 more than expected domains in the giving direction. Each individual who gives more domains than expected averages 0.94 domains above expectation.

Each individual who gets more domains than expected averages 0.79 domains above expectation.

S5: N-gram compared to Importance diversity

*Figure S2* shows the scatter plot of importance diversity by N-gram diversity from the data analyzed in this paper. Half of the people in this sample have no redundant relationships (i.e. N-grams) in these domains (i.e., their importance and N-gram diversity are the same; 126 out of the sample of 252 fall on the 1-1 line). The vertical deviation from the one-to-one line shows how much redundancy people have in their relationship types. As one moves towards the left-hand side of the x-axis, the fewer trading partners one has and the more concentrated trade is with a single partner. As one moves towards the bottom of the y-axis, the more concentrated trade is in a single relationship type.

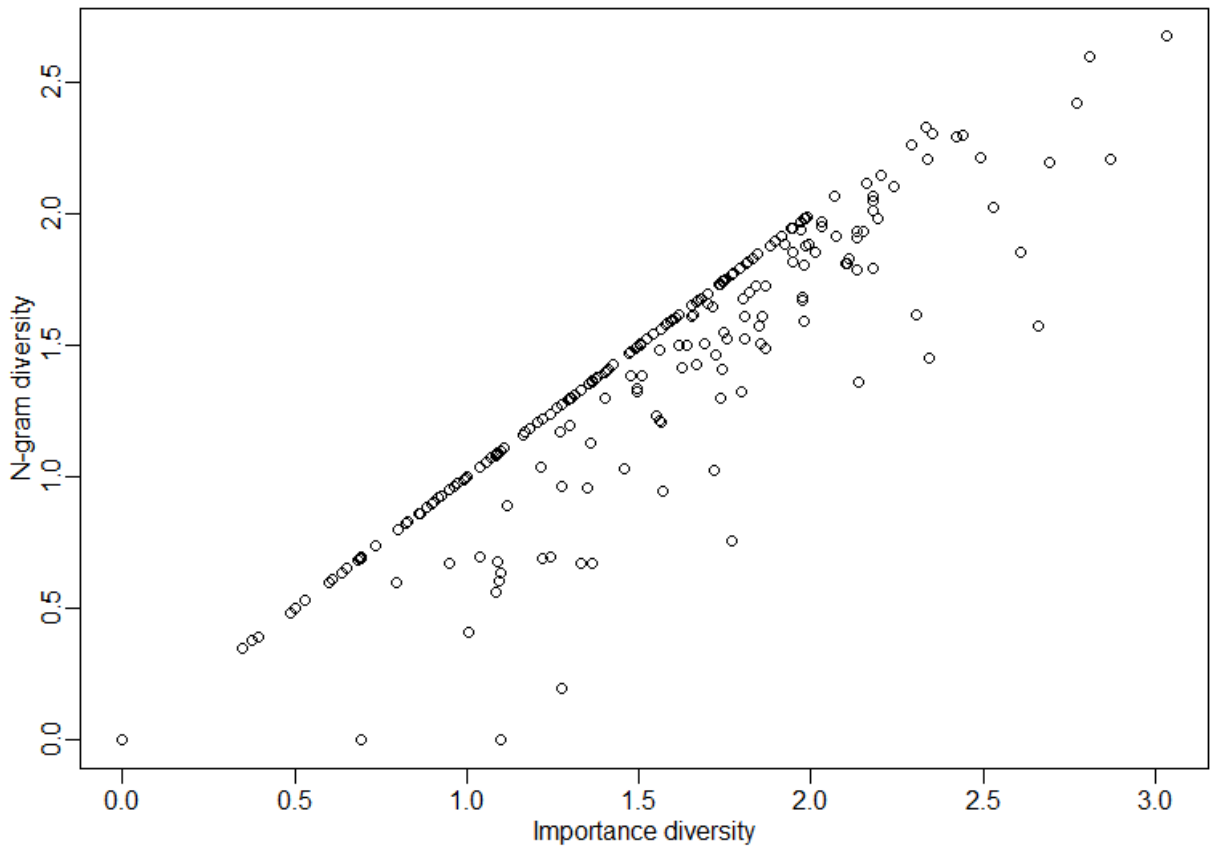

| Outcome | Model ID | No-network-structure | | | | | | Aggregate-layer | | Multiplex | | | | loo_ic |
|---|---|---|---|---|---|---|---|---|---|---|---|---|---|---|
| | | Strength | Error | Domains | Error | Partners | Error | Importance | Error | Ngram | Error | Redundancy | Error | |
| **Skip** | N | -0.01 | 0.01 | 0.05 | 0.02 | 0.01 | 0.02 | ---- | ---- | ---- | ---- | ---- | ---- | 753.19 |
| | N + A | -0.01 | 0.01 | 0.05 | 0.02 | 0.01 | 0.04 | -0.02 | 0.22 | ---- | ---- | ---- | ---- | 755.56 |
| | N + G | -0.01 | 0.01 | 0.04 | 0.02 | -0.03 | 0.03 | ---- | ---- | 0.26 | 0.19 | ---- | ---- | 754.13 |
| | N + M | 0.01 | 0.01 | 0.03 | 0.02 | 0.01 | 0.02 | ---- | ---- | ---- | ---- | -0.71 | 0.32 | 750.05 |
| | N + A + G | 0.01 | 0.01 | 0.03 | 0.03 | 0.01 | 0.04 | -0.69 | 0.37 | 0.71 | 0.31 | ---- | ---- | 752.69 |
| | N + A + M | 0.01 | 0.01 | 0.03 | 0.03 | 0.01 | 0.04 | 0.02 | 0.23 | ---- | ---- | -0.71 | 0.32 | 752.88 |
| **Hungry** | N | -0.01 | 0.01 | 0.07 | 0.02 | -0.01 | 0.02 | ---- | ---- | ---- | ---- | ---- | ---- | 707.55 |
| | N + A | -0.01 | 0.01 | 0.07 | 0.02 | -0.01 | 0.04 | 0.03 | 0.24 | ---- | ---- | ---- | ---- | 709.49 |
| | N + G | -0.01 | 0.01 | 0.07 | 0.02 | -0.02 | 0.03 | ---- | ---- | 0.15 | 0.20 | ---- | ---- | 709.03 |
| | N + M | -0.01 | 0.01 | 0.06 | 0.03 | 0.01 | 0.02 | ---- | ---- | ---- | ---- | -0.30 | 0.32 | 709.15 |
| | N + A + G | -0.01 | 0.01 | 0.06 | 0.03 | -0.01 | 0.04 | -0.26 | 0.37 | 0.32 | 0.31 | ---- | ---- | 711.13 |
| | N + A + M | -0.01 | 0.01 | 0.06 | 0.03 | -0.01 | 0.04 | 0.06 | 0.24 | ---- | ---- | -0.30 | 0.32 | 711.45 |

Table S1: Estimated effect of measures derived from the multiplex network in 6 models of 2 outcomes

|  | Skip | | Hungry | |
| --- | --- | --- | --- | --- |
| Model ID | Mean loo_ic | Std. error | Mean loo_ic | Std. error |
| N | 753.19 | 18.63 | 707.55 | 16.79 |
| N + A | 755.56 | 18.79 | 709.49 | 16.90 |
| N + M | 750.05 | 19.65 | 709.15 | 17.09 |
| N + A + M | 752.88 | 19.89 | 711.45 | 17.23 |
| (N) - (N + A) | -2.36 | 0.52 | -1.94 | 0.45 |
| (N) - (N + M) | 3.15 | 4.90 | -1.60 | 2.31 |
| (N) - (N + A + M) | 0.32 | 4.96 | -3.90 | 2.35 |
| (N + A) - (N + M) | 5.51 | 4.87 | 0.34 | 2.39 |
| (N + A) - (N + A + M) | 2.68 | 4.90 | -1.97 | 2.30 |
| (N + M) - (N + A + M) | -2.83 | 0.62 | -2.30 | 0.67 |

**Table S2: Information criteria for multiple models and their comparisons**